# Fused electrical impedance tomography using an electromagnetically tracked handheld probe for oral cancer margin assessment: reconstruction methods and phantom validation

**Sophie A Lloyd[1*], Ehsan Nasiri[1], Allaire F Doussan[1], Ryan J Halter[1,2] and Ethan K Murphy[1]**

[1] Thayer School of Engineering, Dartmouth College, Hanover, NH 03755, United States of America
[2] Geisel School of Medicine, Dartmouth College, Hanover, NH 03755, United States of America

*Author to whom any correspondence should be addressed.

**E-mail:** Sophie.lloyd.th@dartmouth.edu

## Abstract

*Objective:* Positive margins in oral squamous cell carcinoma (OSCC) resection are common and significantly impact patient outcomes. Electrical impedance tomography (EIT) can be used to intraoperatively distinguish cancer from healthy tissue. The objective of this study was to develop and identify a fused EIT approach to accurately reconstruct a tumor boundary using multiple impedance measurements recorded with a handheld electrical impedance probe. *Approach:* Simulations and noise analyses were used to investigate the best reconstruction method for fused EIT, and a measured phantom experiment was conducted to evaluate the EIT reconstructions. *Main results:* Fused approaches successfully and accurately recovered the conductivity boundary location from multiple probe sites, increasing boundary localization to within 0.5 mm of the true boundary for both simulated and experimental results. Simulations and noise analysis revealed that all reconstruction methods maintain strong classification accuracy with added noise on a half-plane boundary scenario and fused difference EIT is the most robust to all types of added noise, due to the high data redundancy in the reconstruction formulation. *Significance:* The measured multi-site probe experiments represent important steps in the development of an approach that can combine EIT from multiple locations to increase boundary localization accuracy. Overall, this result is a step towards a clinically deployable impedance imaging approach to scanning the entire tumor boundary, which could significantly help to improve surgical outcomes for OSCC.

## 1. Introduction

Over 60,000 new cases of oral squamous cell carcinoma (OSCC) will be diagnosed in the United States this year, the majority of which present at an advanced stage requiring primary surgical resection (Siegel et al., 2025). Intraoperative margin status is among the strongest determinant of five-year survival of all solid tumor cancers affecting both men and women (National Cancer Institute, 2026). In practice, positive-margin rates in oral cancer resection range from 1-44% depending on tumor stage, surgeon experience, and the definition of “positive” (Binahmed et al., 2007; Luryi et al., 2014; Orosco et al., 2018). Frozen section analysis (FSA), the only routinely available intraoperative margin-assessment tool, suffers from sampling bias that limits sensitivity (DiNardo et al., 2000; Pathak et al., 2009), and the ability to reliably map positive-margin sites onto the operative field which restricts successful re-excisions to under 60% of cases (Horwich et al., 2021; Nentwig et al., 2021). These constraints motivate the development of real-time, whole-specimen imaging modalities capable of mapping margin status during surgery.

Electrical impedance spectroscopy (EIS) can differentiate cancerous from healthy oral mucosa (Ching et al., 2010; Lloyd et al., 2025; Murdoch et al., 2014; Sun et al., 2010), but existing approaches provide only point measurements insufficient for comprehensive margin assessment. Hu et al. (2022) developed a handheld electrode array for oral cancer detection; however, its 13 mm sensing area and 4 mm electrode spacing limit the minimum detectable lesion size. One strategy for extending the effective imaging area of a compact electrode array is to acquire impedance measurements at multiple spatial locations and computationally fuse them into a single, larger-area map. Murphy et al. (2025, 2020, 2018) demonstrated a fused electrical impedance tomography (EIT) concept using electrodes on a prostate biopsy probe, showing that combining spatially registered datasets produces accurate conductivity reconstructions exceeding the coverage of any single measurement.

Our group has developed a surgical margin assessment (SMA) probe whose electrode array was first validated for tissue discrimination in prostate cancer (Kossmann et al., 2024), then demonstrated for EIT-based conductivity reconstruction in an ex vivo porcine model (Doussan et al., 2025), and most recently shown to differentiate oral cancer from healthy tissue via EIS in a preliminary clinical trial (Lloyd et al., 2025). In this study, we extend the SMA probe's capability by integrating electromagnetic (EM) tracking and adapting the fused EIT framework for oral cancer margin assessment. We present fusing algorithms, simulated experiments with noise analysis, and phantom validation measurements demonstrating, for the first time, that impedance data recorded from an EM-tracked SMA probe at multiple locations on an agar gels can be fused to produce accurate boundary reconstructions. Three fusion strategies are compared: a fused difference EIT (*fused diffEIT*) approach that incorporate all measurement data into one inverse problem, and two post-reconstruction approaches that fuse together multiple single-site conductivity reconstructions (*Averaged* & *Super-resolution*). Phantom experiments include translations across a conductivity boundary and varied probe orientations, establishing the feasibility of fused difference-EIT for improved spatial coverage in oral cancer margin detection.

## 2. Methods

### 2.1 Measurement system

The measurement system employs a tetrapolar drive-sense configuration, where spatially separated current-injection (II) and voltage-sensing (VV) electrode pairs are used. Tetrapolar measurements minimize contact impedance at the electrode-tissue interface, a consideration particularly important at lower frequencies. The electrode array, based on previously validated designs (Doussan et al., 2025; Kossmann et al., 2024; Lloyd et al., 2025), features a 5 × 5 grid of small, 0.6 mm, voltage-sensing electrodes surrounded by eight larger current-injection electrodes (see Figure 1(b)). The larger peripheral electrodes increase surface area, lower contact impedance, and enhance the signal-to-noise ratio. Due to the 32-channel constraint of the acquisition hardware (EIT32, Sciospec, Bennewitz, Germany) the central electrode is omitted, leaving 24 active voltage-measurement electrodes. This configuration enables 28 unique II combinations and 276 distinct VV pairs, producing 7728 unique IIVV permutations. For each measurement frame, impedance spectra are recorded across all IIVV combinations at 31 logarithmically spaced frequencies spanning 100 Hz to 100 kHz. Acquisition of all 28 injection patterns with simultaneous voltage recordings requires approximately 3 seconds per complete spectral frame.

### 2.2 Instrument Tracking

The 3D-printed housing of the impedance probe was modified to accommodate an electromagnetic (EM) tracking sensor (Model 130 6DOF coil, Aurora, Northern Digital Inc., Ontario, Canada), which was rigidly affixed using medical-grade epoxy (LOCTITE® 4013™ Prism™ Medical Device Adhesive) approximately 3.5 mm behind the center of the electrode array. The Tabletop NDI Aurora system reports a positional accuracy of 0.7 mm root mean square (RMS) and an orientational accuracy of 0.3° RMS with the field generator used in this study (Aurora, Northern Digital Inc., Ontario, Canada). Three pinholes were incorporated into the cap of the probe to serve as orientation registration landmarks aligned with the center of the electrode array, as shown in Figure 1(a).

An EM-tracked stylus, calibrated via pivot calibration (calibration error < 0.1 mm) before each use, was employed for all fiducial localization tasks described below.

To determine the rigid-body transformation from the EM sensor to the center of the PCB-based electrode array ($T_{pcb\leftarrow coil}$), a custom registration jig was designed and 3D-printed using clear resin (Formlabs, MA, USA). The jig contained six fiducial divots into which the probe tip was sequentially inserted in known poses, as shown in Figure 1(b). The ground-truth positions and orientations of the fiducial divots were obtained from a computed tomography (CT) scan of the jig. The jig was registered from CT to EM space using nine fiducial points (six divot centers and three additional points distributed across three orthogonal planes of the jig housing) yielding a fiducial registration error (FRE) of 0.75 mm root mean square error (RMSE). The transformation $T_{pcb\leftarrow coil}$ was then estimated using nonlinear least-squares optimization over both translational and rotational residuals at each of the six fiducial poses. The resulting target registration error, evaluated in CT space, was 0.57 mm RMSE (translational) and 2.2° mean rotational error.

Housings for the gel phantoms were custom designed and fabricated with 3D-printed clear resin, shown in Figure 1(c). Eight fiducial divots were incorporated into each phantom housing (four on the top corners and four distributed across the four perpendicular side faces) to enable registration of the phantom to the EM tracker coordinate system. Each fiducial was localized using the EM-tracked stylus, and the corresponding ground-truth positions were extracted from CT. Phantom registration yielded an FRE of 0.55 mm RMSE, computed across all fiducial points from three repeated registration trials.

### 2.3 Phantom Design and Data Acquisition

The phantom consisted of a rectangular box divided at the center by a straight boundary, with one half filled with agar gel (Sigma Aldrich; conductivity 0.44 S/m, approximating muscle) and the other half filled with saline (conductivity 0.1 S/m, more conductive than fat) (Gabriel, 1996). Saline was added immediately before data collection, and all measurements were completed within 10 minutes to minimize ion diffusion across the gel–saline interface. Phantom registration was performed four times using fiducial locations from the design file as ground truth, yielding a mean registration RMSE of 0.60 mm.

Forty-two measurements were collected across the phantom with hand-held positioning. The first measurement was taken at the center of the saline region, as far as possible from both the phantom walls and the gel boundary, and served as the homogeneous background reference for difference reconstructions. The final measurement was taken at the center of the gel region to provide a pure-gel reference. The remaining 40 measurements were acquired at approximately 1 mm intervals across the boundary, traversing from saline to gel and returning to the saline side.

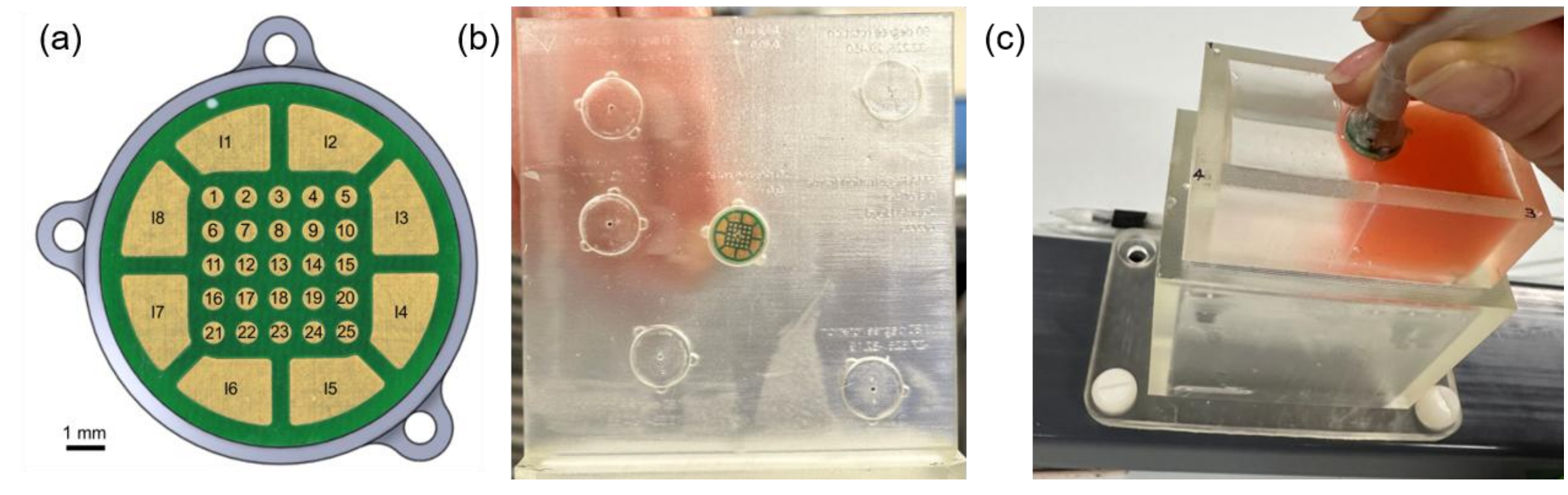

**Figure 1. Registration and testing setup with housing for half-plane agar phantom and probe electrode array geometry.** (a) SMA probe electrode array in three pin housing with current drive electrodes labelled with "I" and voltage electrodes numbered from 1 to 25, (b) Custom jig for EM coil to electrode registration, and (c) EM-tracked probe taking a measurement on half plane gel phantom housing.

**2.4 Forward Problem**

The forward problem is governed by the generalized Laplace equation, $\nabla \cdot (\sigma \nabla u) = 0$, where $\sigma$ represents the conductivity and $u$ denotes the electric potential within the domain. The complete electrode model (CEM) is employed to accurately represent the electrode–domain interface (Vauhkonen et al., 1999). Given prescribed electrode currents and a known conductivity distribution, the forward problem determines the electric potential distribution throughout the domain and the resulting electrode voltages. The CEM boundary conditions stipulate that: (1) current enters and exits the domain exclusively through the electrodes; (2) the potential on each electrode is governed by a contact impedance condition, permitting spatial variation of the potential beneath the electrode surface; (3) the integral of the current density over each electrode equals the applied current; and (4) zero current flux is imposed on all boundary surfaces not covered by electrodes (Murphy et al., 2017; Vauhkonen et al., 1999). Because the impedance probe operates on an open domain rather than an enclosed one, the computational domain is extended sufficiently such that the current density is negligible at the outer boundaries, effectively approximating an unbounded medium. The three-dimensional forward problem is solved using the finite element method (FEM) with linear basis functions.

**2.5 Inverse Methods**

A Gauss-Newton algorithm with Laplace-smoothing Tikhonov regularization is utilized for solving the inverse problem. At each probe position, standard electrical impedance tomography (EIT) data, referred to as a frame of data, is collected. This research considers fused-data EIT, which combines frames from multiple probe positions within a wider open domain. The methodology employs a method similar to the approach used in Murphy et al. (2018), with the aim to minimize the following objective function:

$$E(\delta\sigma) = \sum_{n=1}^{N_s} \|J_n \delta\sigma - \Delta v_n\|_2^2 + \lambda \|L(\sigma_0 + \delta\sigma)\|_2^2, \tag{1}$$

where $N_s$ is the number of frames of data fused together, $J_n$ is the Jacobian corresponding to the nth state, $\Delta \boldsymbol{v_n}$ is the voltage differences of the nth state, $\lambda$ is the Tikhonov factor, $L$ is the regularization matrix, $\sigma_0$ is the initial guess of the conductivity distribution, and $\delta\sigma$ is the perturbation being solved for at each iteration. For each iteration, $J_n$ is size $N_P \times N_C$ and $\Delta v_n$ is size $N_P \times 1$, where $N_P$ is the number of tetrapolar measurement patterns (7728) and $N_C$ is the number of coarse inverse voxels. To account for multiple measurement sites, Equation 1 can be rewritten equivalently as:

$$E(\delta\sigma) = \|J_{FD} \delta\sigma - \Delta v_{FD}\|_2^2 + \lambda \|L(\sigma_0 + \delta\sigma)\|_2^2. \tag{2}$$

$J_{FD}$ and $\Delta v_{FD}$ are concatenated matrices and vectors from all measurements of sizes $N_s * N_P \times N_C$ and $N_s * N_P \times 1$, respectively. From this formulation, the standard updating formula can be derived as:

$$\delta\tilde{\sigma} = (J_{FD}^T J_{FD} + \lambda L^T L)^{-1} \, (J_{FD}^T \Delta v_{FD} - \lambda L^T L \sigma_0). \tag{3}$$

This study uses difference reconstructions, making $\Delta v_{FD}$ the difference between a test measurement and a single reference measurement of homogeneous conductivity, taken at the beginning of the experiment from the center of the saline half. The fully concatenated formulation (3) is solved directly for *fused diffEIT*, while single probe site reconstructions (*Average* & *Super-resolution*) use a similar formulation, but do not concatenate Jacobians or voltage differences as each measurement site is reconstructed separately (i.e., (1) is solved for each $n$ independently).

Regardless of fusion approach, a dual-mesh method (Borsic et al., 2010) is employed, in which the Jacobian is computed on a fine forward mesh for accurate electric potential estimations while conductivity is estimated on a coarser inverse mesh. The fine mesh, comprising approximately 130k nodes, was generated using gmsh (Geuzaine and Remacle, 2009). To model the open domain, the mesh was extended 80 mm in the x- and y-directions from the probe center and 40 mm axially from the face of the probe, along the z-direction. All simulations were performed using The New Dartmouth Reconstructor Matlab (NDRM) toolbox (Borsic et al., 2008).

Single-site reconstructions were solved on a coarse inverse mesh comprising 1,426 nodes distributed across five axially spaced z-layers extending 1.2 mm into the domain. Each layer contains a variable number and size of voxels; the shallowest layer has the highest density (360 voxels), spanning ±4.7 mm from the center in both the x and y directions, while the deepest layer contains 153 voxels. The coarse mesh extends just beyond the outer current-drive electrodes, and a mega-node approach (single conductivity value) is used for regions outside the high-sensitivity area beneath the electrode array (Murphy et al., 2016). Reconstructed conductivity was restricted to regions meeting a sensitivity threshold of 1/50 of the peak Jacobian value, previously determined to be optimal for reconstructions with this electrode array (Kossmann et al., 2024).

The extent of the fused coarse grid was determined by applying the known rigid-body transformations from each measurement position to the single-site coarse grid and computing the convex hull of all transformed grids. The convex hull was discretized as a rectilinear mesh with 0.3 mm spacing, yielding a fused coarse grid of 5,819 voxels across five depths, spanning 13.3 mm in x and 9.7 mm in y at the shallowest z-layer. Both the sensitivity threshold and node spacing were selected after systematic evaluation of multiple candidate values. Heuristic tikhonov regularization parameters of $\lambda=1\times10^5$ and $\lambda=1\times10^7$ were used for single-site and *fused diffEIT* reconstructions, respectively.

### 2.6 Post Reconstruction Fusing Methods

Two methods were explored to fuse multiple single-site reconstructions after they were computed independently. These methods have the potential to reduce computational costs in the reconstruction process and are less limited by the number of measurements that can be fused as compared to the *fused diffEIT* method.

#### 2.6.1 Super-Resolution Fusion

The first is a robust multi-image super-resolution (SR) method adapted from Farsiu, et al. (2006). The SR framework defines a forward model of the imaging channel, incorporates prior information such as known measurement positions to regularize the ill-posed inverse problem, and fuses the information from multiple images in a robust and computationally efficient manner. The forward model is:

$$\delta\tilde{\sigma}(n) = D(n)F(n)X \quad\quad n = 1, \ldots, N_s\ , \tag{4}$$

where $X$ is the high-resolution (HR) image to be estimated, $\delta\tilde{\sigma}(n)$ is the $n^{th}$ low resolution (LR) frame, $F(n)$ is the geometric motion operator between the HR frame to the $n^{th}$ LR frame, $D(n)$ is the decimation operator, and $N_s$ is the number of LR frames. The HR image $X$ has $r$ times more pixels than each $\delta\tilde{\sigma}(n)$ in both x and y directions. A value of *10* was used for *r* in this work. The blur operator *H(n)* and additive noise term from the original formulation of Farsiu et al, (2006) were omitted from our implementation, as these components were not applicable to our electrical impedance reconstruction data. The motion operator was derived from known probe displacements.

The inverse problem is solved using the robust cost function of Farsiu et al, (Farsiu et al., 2006), which combines an L1-norm data fidelity term with a Bilateral Total Variation (BTV) regularizer. The L1 norm provides robustness against outliers, while the BTV regularization preserves edge sharpness and suppresses artifacts. Minimization is performed via iterative steepest descent:

$$\hat{X}^{n+1} = \hat{X}^n - \beta \left\{ \sum_{n=1}^{N_s} F^T(n) D^T(n) sign(D(n)F(n)\hat{X}^n - \delta\tilde{\sigma}(n)) + \gamma \sum_{l,m=-P}^{P} \alpha^{|m|+|l|} \left[ I - S_y^{-m} S_x^{-l} \right] sign(\hat{X}^n - S_x^l S_y^m \hat{X}^n) \right\} \quad (5)$$

where $\beta$ is the step size, $\gamma$ is the regularization parameter controlling the balance between data fidelity and smoothness, and α (0 < α < 1) applies a spatially decaying weight to the regularization terms. The operators $S_x^l$ and $S_y^m$ shift the image by l and m pixels in the horizontal and vertical directions, respectively, while $S_x^{-l}$ and $S_y^{-m}$ represent shifts in the opposite directions. The parameter P determines the spatial extent over which the BTV derivatives are computed. Parameters were determined using manual tuning for both simulated and experimental reconstructions. The SR algorithm was run for a maximum of 100 iterations with parameters: $\beta = 50, \gamma = 0.01, P = 3.0, \alpha = 0.5$ for simulated data. Experimental data used parameters: $\beta = 70, \gamma = 0.07, P = 3.0, \alpha = 0.5$ with a maximum of 100 iterations.

**2.6.2 Spatially Averaged Fusion**

The second fusion approach employs spatial averaging (Avg) based on interpolation and registration of individual LR reconstruction frames. Each LR conductivity reconstruction frame $\delta\tilde{\sigma}(n)$ is spatially registered according to its known position shift, upsampled to the HR grid through 2D interpolation, and the final HR image is obtained by computing the pixel-wise mean across all $N_s$ registered and interpolated frames:

$$\hat{X} = \left(\frac{1}{N_s}\right) \sum_{n=1}^{N_s} \Phi\{F(n)\delta\tilde{\sigma}(n)\}, \quad (6)$$

where $\Phi\{\cdot\}$ denotes the 2D interpolation operator that registers and upsamples the $n^{th}$ LR frame to the HR grid after application of the motion operator, $F(n)$. Nearest-neighbor interpolation was performed and boundary voxels were set to the prior edge value when frames did not fully overlap on the HR grid.

For both the SR and averaged approaches, the cropped single-site reconstructions (10×10 voxels) were used as LR inputs. An upsampling factor of r=10 was empirically selected, yielding HR output images of 100×100 voxels.

**2.7 Noise Analysis**

To simulate experimental data conditions, noise was incorporated within the simulations in three ways. First, Gaussian noise, scaled by the reference saline simulation, was added to the simulated impedance data at different variance levels. Gaussian noise was added at twelve variance levels spanning 1 to 100% of the reference saline impedance values. Second, up to 12 randomly selected electrodes were designated as having poor contact; consequently, all current-injection voltage measurements (IIVVs) associated with these electrodes were filtered out. Both voltage-sensing and current-driving electrodes were eligible for elimination. Third, errors from the EM tracking sensor were simulated by adding Gaussian noise at various levels to both the x/y position and orientation of the probe measurements (0.1, 0.5, 1.0 mm positional error; 1%, 5%, 10% error in orientation). For the *fused diffEIT* reconstructions, the fused coarse grids and Jacobians were constructed with the noisy positions and orientations, but the ideal simulation impedance data was used for reconstruction, thus simulating a mismatch between where we believed the measurements to be taken and where they were actually measuring. For SR and Avg, the ideal simulated reconstructions were shifted by the noisy positions before fusing occurred.

**2.8 Edge Spread Function Analysis**

Spatial resolution and boundary localization accuracy of each reconstruction were quantified from the centerline conductivity profile across the ground truth conductivity step (i.e., gel-to-saline transition). For each reconstruction, the centerline profile was extracted along the row passing through the ground-truth boundary, yielding a one-dimensional edge spread function (ESF) (Samei et al., 1998). Each ESF was normalized between plateau values. The steepest descent of a smoothed gradient identified an initial edge estimate and the high and low plateau values were taken as the median of points lying more than 0.8 mm to either side of this estimate. The normalized ESF was then fit, in a least-squares sense, to a complementary error function model:

$$ESF(x) = \frac{1}{2}\left[\mathrm{erfc}\left(\frac{x - x_0}{w\sqrt{2}}\right)\right], \tag{7}$$

with edge center $x_0$ and edge-width parameter $w$ (the 1/e half-width of the underlying Gaussian line spread function (LSF)) as free parameters. From the fit we derived four metrics: (i) the edge-localization error, $\Delta x = x_0 - x_{\mathrm{truth}}$, quantifying the displacement of the fitted edge from the known boundary position; (ii) the full-width at half-maximum (FWHM) of the analytic LSF, computed as $\mathrm{FWHM} = 2\sqrt{2\ln 2}\ w$, providing a measure of boundary blur; (iii) the 10-90% edge width, computed by inverse interpolation of the fitted ESF and serving as a complementary, monotonicity-based resolution metric less sensitive to the assumed Gaussian LSF shape; and (iv) the fit residual ($R^2$ of the erf model relative to the normalized data), used to flag reconstructions whose ESF departed substantially from an erf profile. For the single-site reconstructions, all four metrics were computed per source position and reported as the mean and standard deviation across the $N = 12$ positions. The same metrics were computed on each fused reconstruction. Resolution metrics ($w$, FWHM, 10-90% width) were used to compare boundary blur between modalities, while $\Delta x$ was used to compare boundary-localization accuracy.

## 3. Results

Only measurements taken within 2 mm of the phantom centerline were included in the fused reconstructions to maintain boundaries within the most sensitive region of the probe. Twelve measurements met this condition, with shifts ranging from those centered 1.87 mm to the right of center and 1.78 mm to the left, and six measurements on each side of the center position. All reconstructions were cropped to the central ± 2.3 mm of the inner electrode array and ensure comparison over a common region of interest.

As currently implemented, *fused diffEIT* requires a common set of IIVV patterns across all measurement sites; consequently, any pattern filtered from one measurement was removed from all. While this resulted in ~25% total data loss in the experimental study, ~69k IIVV patterns from the total of >92k IIVVs from the 12 probe positions ultimately used were still available for assessing reconstruction performance on the gel–saline phantoms.

### 3.1 Simulation

Figure 2 shows the simulated reconstructions at depths: -1.2, -0.91, -0.65, and -0.39 mm on method-appropriate meshes (10x10 single, 100x100 SR and Avg, 14x14 *fused diffEIT*). All methods produced clear contrast between the gel (0.44 S/m) and saline (0.1 S/m) regions.

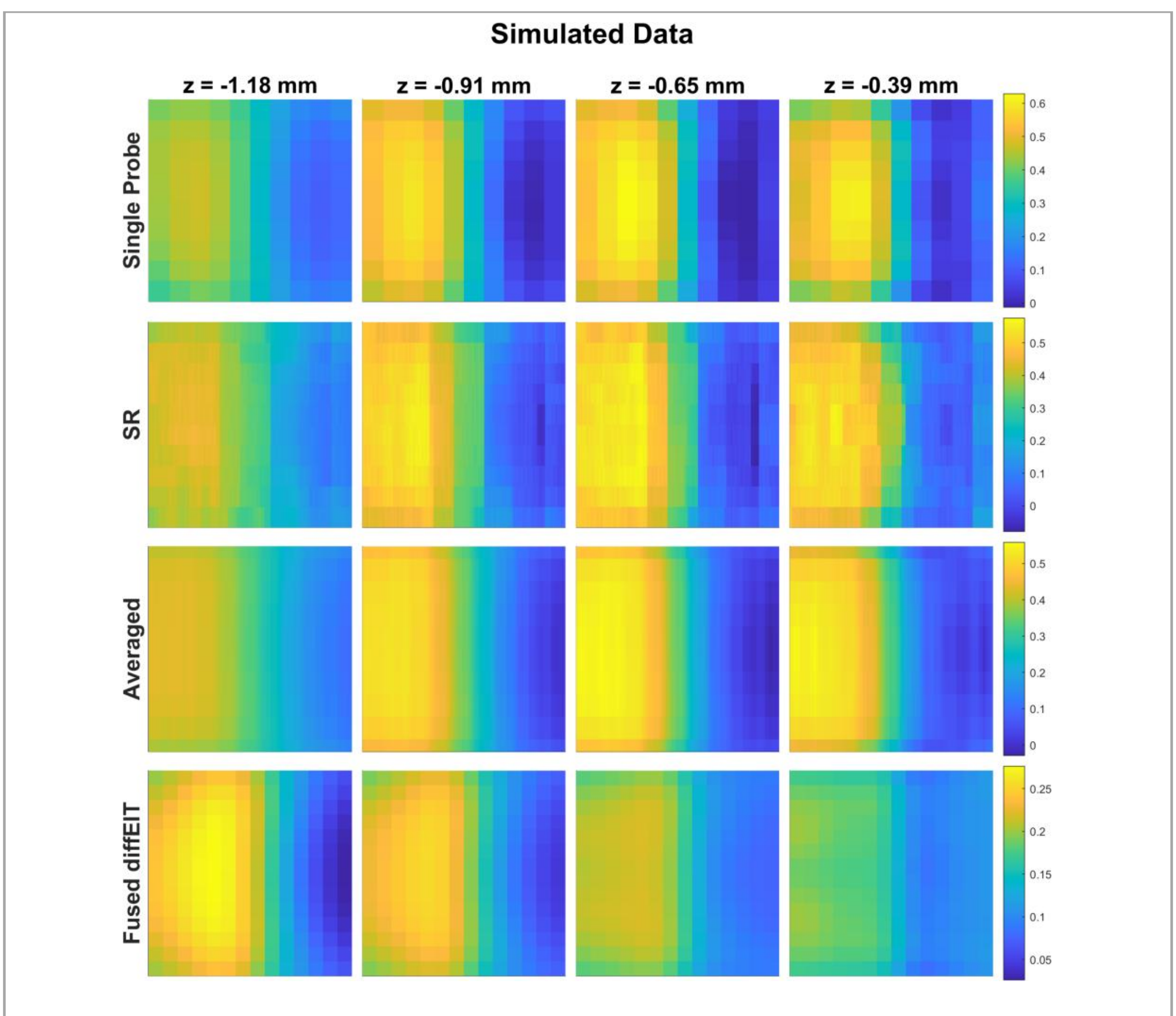

**Figure 2. Simulation reconstructions at depth for the four reconstruction methods**. The conductivity range (S/m) for each method is shown by the colorbar on the right of each row.

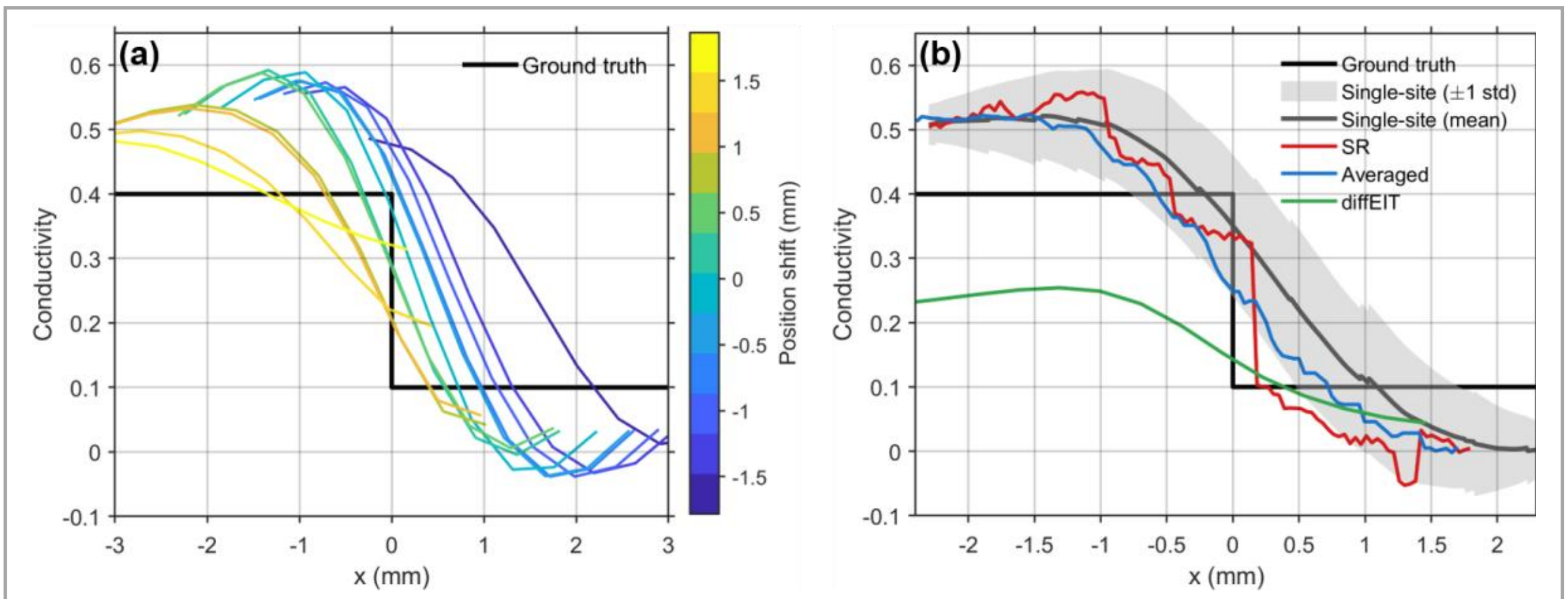

**Figure 3. Edge spread functions (ESFs) of simulated single-site and fused EIT reconstructions.** (a) Conductivity curves from 12 single-site reconstructions across positions spanning -1.5 to 1.8 mm from the boundary, aligned on phantom frame (b) Cross-method comparison: mean ± 1 std of the single-site ESFs (gray), overlaid with the ESFs of the super-resolution, averaged, and fused reconstructions. The black line indicates the ground-truth conductivity step.

**Table 1.** Edge-reconstruction metrics quantifying edge-localization accuracy and quality for all four reconstruction modalities using simulated and experimental data.

| Metric | Data type | Single Site (mean ± std) | SR | Averaged | Fused diffEIT |
|---|---|---|---|---|---|
| Δx (mm) | Sim | 0.59 ± 0.48 | 0.23 | 0.31 | 0.27 |
| | Exp | 0.76 ± 0.50 | 0.24 | 0.14 | 0.34 |
| FWHM (mm) | Sim | 1.28 ± 0.11 | 1.15 | 1.61 | 1.26 |
| | Exp | 1.21 ± 0.07 | 1.50 | 3.39 | 2.18 |
| 10-90% (mm) | Sim | 0.76 ± 0.51 | 1.25 | 1.75 | 1.37 |
| | Exp | 1.31 ± 0.08 | 1.63 | 3.68 | 2.34 |
| $R^2$ | Sim | 0.99 ± 0.00 | 0.98 | 1.0 | 0.99 |
| | Exp | 0.99 ± 0.00 | 0.99 | 1.0 | 1.0 |

[a] Δx – edge-localization error, FWHM – full width at half-maximum, Sim – simulated data as input, Exp – experimental data as input.

The edge spread functions (ESFs) extracted along the midline at z=-0.91 mm from all 12 single site positions are shown in Figure 3(a), with edge-localization metrics for every modality summarized in Table 1. Ideally, reconstruction methods would have close to zero localization error (Δx), limited blurring of the conductivity boundary as quantified with FWHM and 10-90% rise width, and a good fit with the ESF demonstrated by $R^2$ near 1. As the probe shifted toward the high-conductivity half, the reconstructed conductivity range decreased and the central region was estimated above 0.3 S/m. Edge-localization error grew with probe offset: ESF curves plotted in the phantom frame should align at 0, but the most offset single-site boundaries appear up to 1.5 mm from ground truth.

All ESF fits had $R^2$ >0.98. Super-resolution produced the most accurate boundary (0.23 mm error) and the lowest FWHM, indicating the least blur. The single-site mean had the largest error (0.59 ± 0.48mm), but the narrowest 10-90% transition. This may arise because the single-site voxels are an order of magnitude larger than those in the SR reconstruction, so the boundary is typically resolved across only one or two voxels, producing a sharp but spatially coarse transition. *Fused diffEIT* achieved localization accuracy (0.27 mm) comparable to SR, and outperformed the averaged reconstruction on both FWHM and 10-90% spread. However, it yielded the smallest reconstructed conductivity range of all modalities, suggesting amplitude attenuation from the regularization needed to combine all 12 measurements in the inverse problem.

### 3.2 Noise in Simulations

#### 3.2.1 White Noise

Simulations model a configuration in which electrodes are in direct contact with the phantom, where the two regions exhibit a 4:1 conductivity contrast (0.44 versus 0.1 S/m between agar and saline). Single-site metrics were averaged across all twelve measurement locations; both SR and averaged reconstructions used these single-site results as inputs at each noise level.

To verify that robustness across noise levels reflects genuine signal preservation rather than reconstruction artifact, two IIVV patterns with sensitivity localized to the agar and saline halves, respectively, were examined directly. Their noise-free gel-to-saline impedance ratio was 3.99 while the 100% added noise ratio was 2.83, confirming that the intrinsic conductivity difference between the two regions produces an impedance contrast large enough to remain well above the noise floor even at extreme variance levels.

Figure 4(a-c) shows edge-localization error, FWHM and 10-90% width as functions of noise level. Boundary localization accuracy was preserved across noise levels for all modalities. Single-site reconstructions were most noise-sensitive, with FWHM and 10-90% width fluctuating >0.6mm above 60% noise alongside increased variability. *SR* and *averaged* reconstructions showed >0.3 mm blur changes beginning at 50% noise, while *fused diffEIT* was the most robust, with a maximum FWHM and 10-90% change of 0.3mm even at 100% noise.

**3.2.2 Electrode Removal**

To emulate poor electrode contact at one corner of the array, twelve electrodes were removed sequentially, beginning with two voltage electrodes in a single corner (see Fig. 1) and proceeding through the order [voltage (v)1, v2, current (I)1, v6, v7, I8, v3, v8, I2, I7, v4, v11]. Removing two voltage electrodes eliminates ~8% of the 7,728 IIVV patterns; removing one current-drive electrode discards ~25%; the combination eliminates 37%. Removing six voltage and four current electrodes prunes 91%, leaving 720 usable patterns.

Figure 4(d-f) shows the impact of pattern loss. All four methods retained accurate edge-localization even with 90% pattern loss, though all boundaries shifted leftward by 0.25 – 1 mm toward the corner of removed electrodes. SR exhibited the largest blur degradation, with FWHM and 10-90% width increasing by 0.72 mm and 0.8 mm, respectively, at 90% loss; single-site, averaged, and diff-EIT methods increased by 0.55-0.65 mm. *Fused diffEIT* was the most stable, maintaining near-baseline metrics up to 70% pattern removal. This robustness arises because aggregation across the twelve shifted measurements retains 8,640 IIVV patterns even after 90% of individual patterns are discarded from each individual measurement.

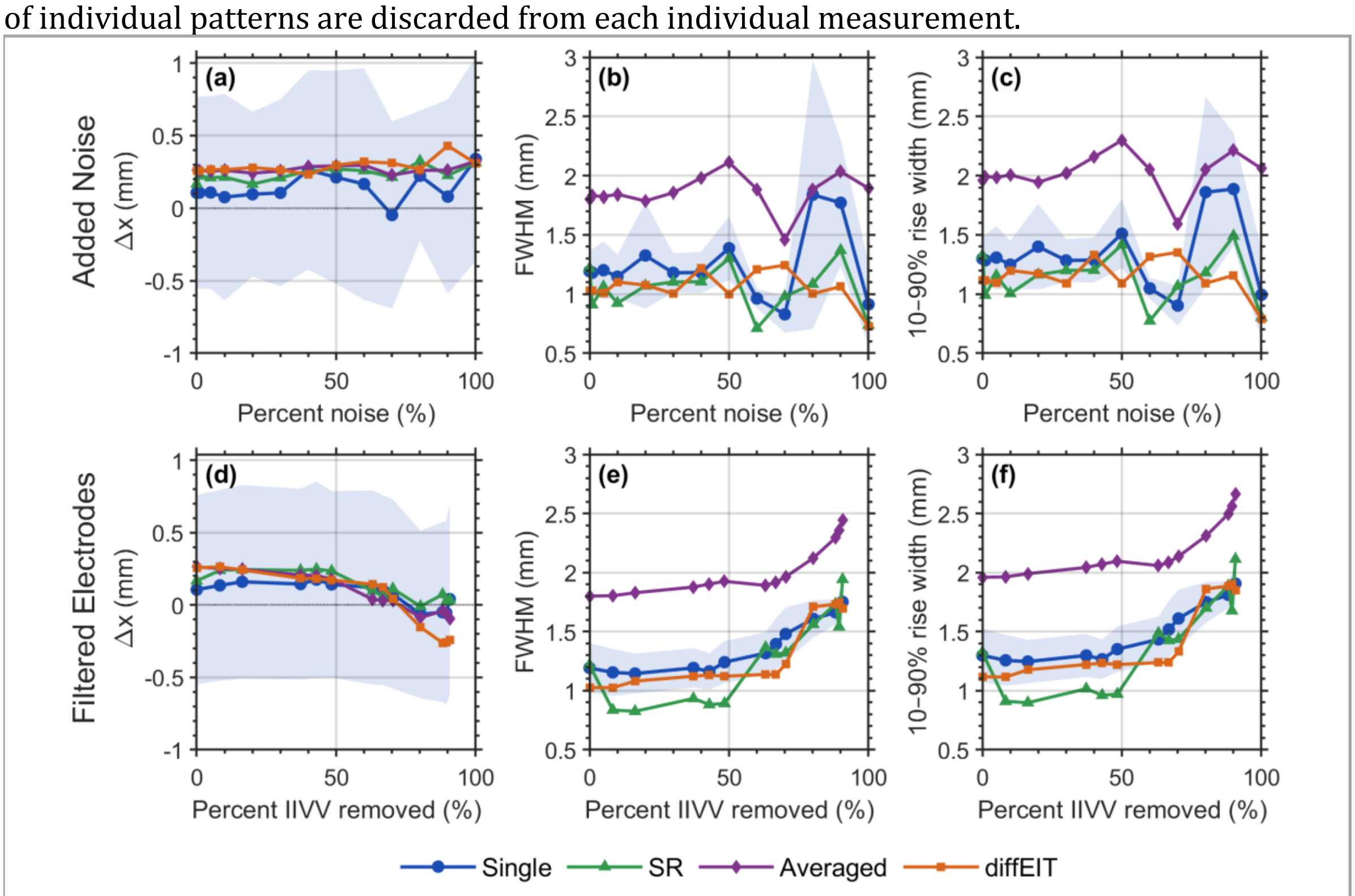

**Figure 4. Edge-reconstruction quality degrades with increasing measurement noise (top row) and electrode filtering (bottom row) across all four modalities.** Three metrics quantify edge recovery as a function of stressor level: edge-localization error Δx (a,d); full-width at half-maximum (b,e); and 10-90% rise width (c,f). Top row: increasing measurement noise. Bottom row: increasing fraction of filtered electrodes. Single-site reconstructions (blue) are shown as mean ± 1 SD across 12 edge positions (shaded band); super-resolution (SR, green), averaged (purple), and fused diffEIT (orange) reconstructions yield a single trace per stressor level.

### 3.2.3 EM Localization Error

Positional and rotational localization errors were each tested at three levels: positional variances of 0.1 mm (ideal tracking error), 0.5 mm (typical experimental error), and 1 mm (maximum allowable error); rotational errors of 1%, 5%, and 10%. All errors were drawn from zero-mean Gaussian distributions with variances scaled to the prescribed level.

*Fused diffEIT* reconstructions were essentially unaffected by any error combination: edge localization, FWHM, and 10-90% all changed by less than 0.01. For SR and averaged reconstructions, localization error and blur generally increased with added noise. SR and averaged yielded the highest localization errors with 0.5 mm positional variance and 10 degrees rotational variance. Rotational variation increased the blur of the SR reconstructions the most, as the FWHM increased from 1.22 to 1.68 and 10-90% increased from 1.33 to 1.83 with 0.1 mm positional error and 10 degrees rotational variance. Averaged reconstructions were more impacted by high positional error, as the highest positional error nearly doubled the FWHM (1.8 to 3.3) and 10-90% (1.96 to 3.04).

### 3.3 Experimental Reconstructions

Figure 5 and 6 show results of 10kHz experimental data, where achieve high SNR with the Sciospec EIT 32. Compared with the simulated results, the experimental reconstructions exhibited greater intra-region inhomogeneity in both the agar and saline regions, particularly at shallow depths. As shown in Figure 5, *fused diffEIT* produced reconstructions with markedly narrower conductivity range (0.05 to 0.2 S/m) than other methods (0.05 to 0.5 S/m). Notably, the *fused diffEIT* conductivity range increased with reconstruction depth, while the range decreased with depth for all three other methods.

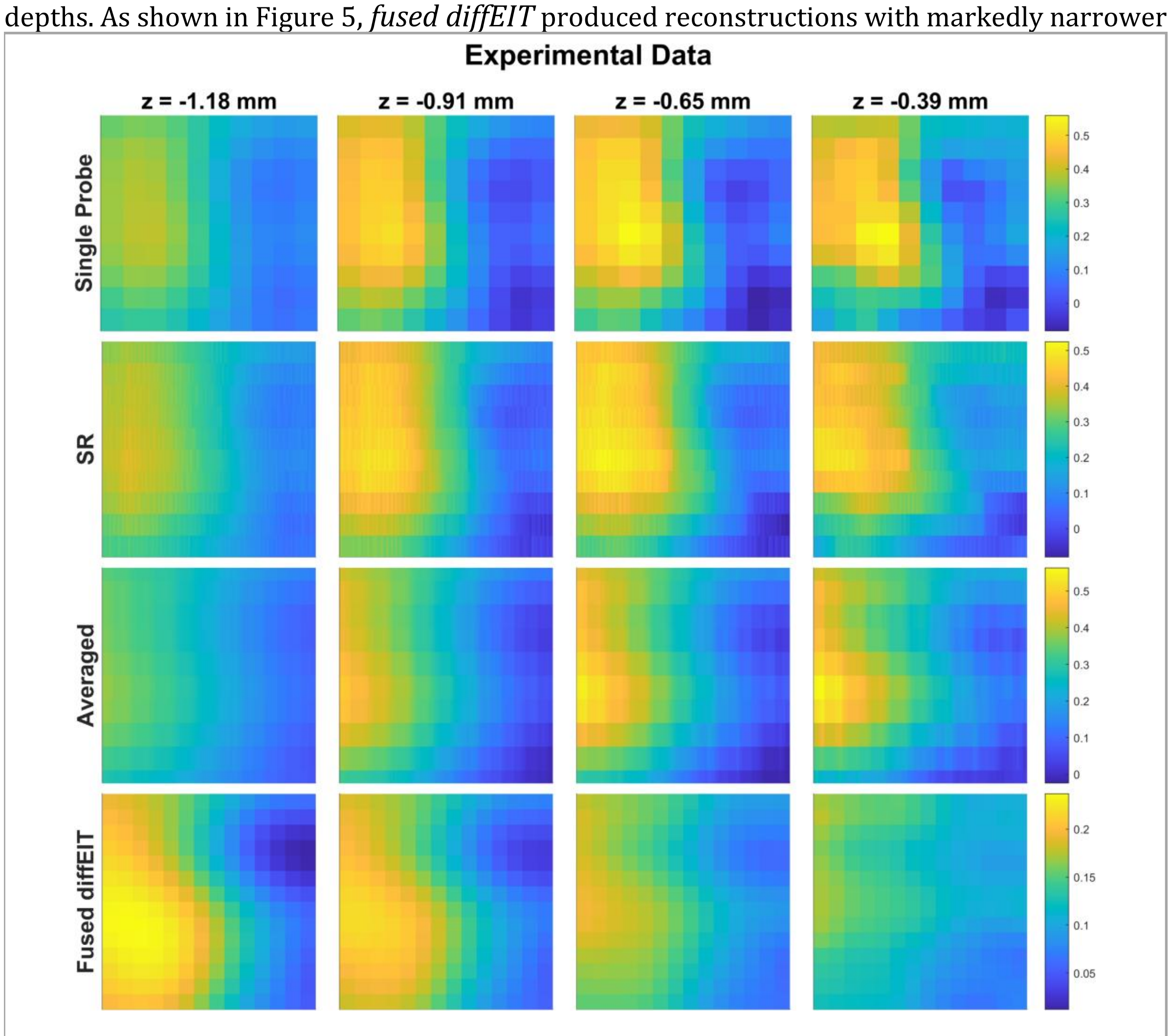

**Figure 5. Experimental reconstructions shown at 10kHz for all methods.** The colorbar indicating the range of conductivity change (S/m) is shown for each method individually.

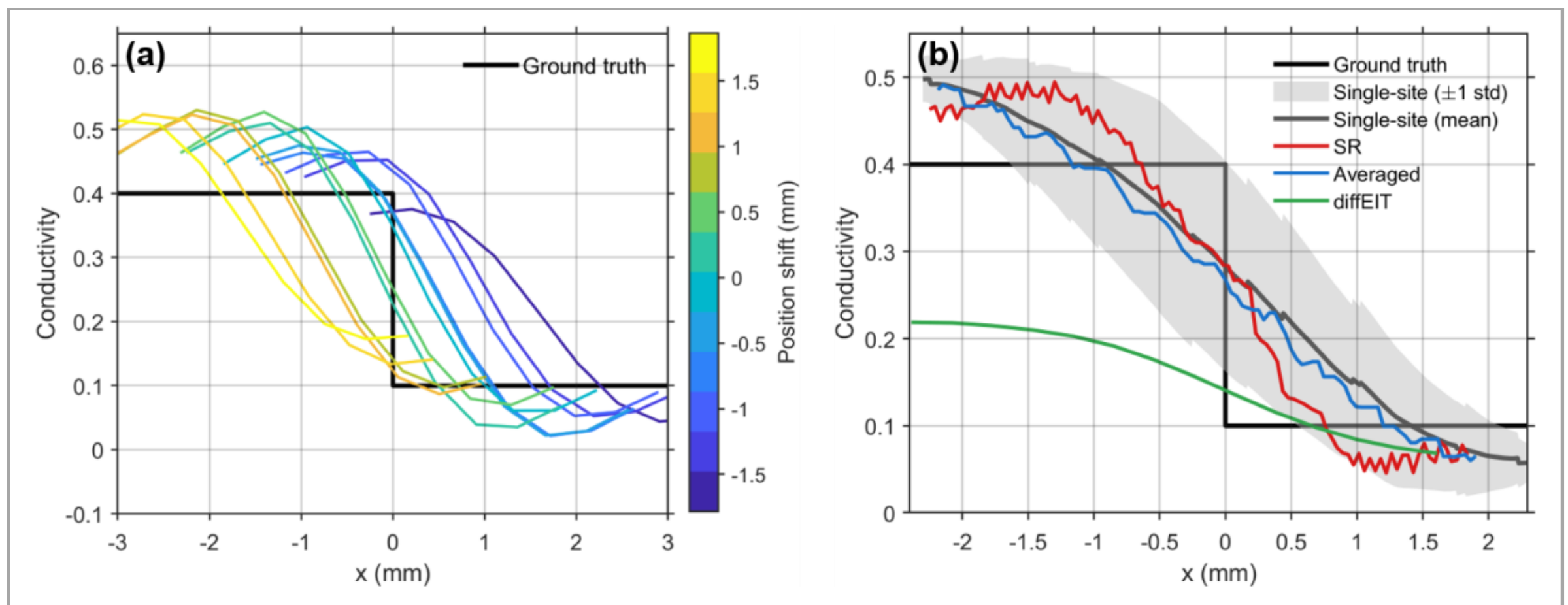

**Figure 6. Edge spread functions (ESFs) of experimental reconstructions.** (a) Center line conductivity curves from 12 single-site reconstructions across positions spanning -1.5 to 1.8 mm from the central boundary (b) Cross-method comparison: mean ± 1 std envelope of the single-site ESFs (gray), aligned on the ground truth frame, overlaid with the ESFs of the super-resolution, averaged, and fused reconstructions. The black line indicates the ground-truth conductivity step.

The single-site experimental reconstructions reproduced the simulation trend in which the reconstructed conductivity range narrows as the probe is shifted further from the boundary. Boundary localization error for single-site reconstructions was 0.76 ± 0.50 mm, approximately 0.2 mm greater than the simulated value (Table 1). Figure 6(b) shows that single-site reconstructions exhibited higher variance in both reconstructed conductivity and boundary position across probe positions than in simulation.

A blur-localization trend trade-off emerged across methods. Single-site reconstructions achieved the lowest blur (FWHM = 1.21 mm, 10-90% width = 1.31 mm) but the largest localization error. Averaged reconstructions achieved the most accurate boundary localization (0.14 mm) but the highest blur (FWHM = 3.7 mm). SR and *fused diffEIT* occupied intermediate positions: SR with FWHM = 1.5 mm and localization error 0.24 mm; *fused diffEIT* with FWHM = 2.2 mm and localization error of 0.34 mm.

## 4. Discussion

In this study, a previously developed SMA probe was adapted for EM-tracking and used with three fusion approaches: *fused diffEIT*, *SR*, and averaged reconstructions. EIT methods were developed and validated against simulated and experimental phantom data. This work demonstrates the potential to convert a point-wise diagnostic tool into a spatially resolved margin-mapping system deployable intraoperatively, addressing the lack of comprehensive registered coverage in FSA.

All methods achieved accurate boundary localization, with errors under 0.4 mm for the fused methods and mean errors below 1 mm for single-site reconstructions. However, no single method optimized all metrics: single-site reconstructions produced the sharpest boundaries within their limited 4.6 × 4.6 mm field of view; averaged reconstructions yielded the most accurate boundary localization in experimental data but the most diffuse transitions; and *SR* and *fused diffEIT* balanced these properties, with *fused diffEIT* additionally offering the strongest noise robustness. This trade-off implies that the optimal fusion method depends on the clinical priority between edge sharpness, localization accuracy or robustness.

Single-site reconstructions avoid the smoothing introduced by data fusion and were evaluated on a region optimally centered within the probe's region of highest sensitivity. This method is the most computationally efficient: all 12 reconstructions were computed in 3.97 seconds with precomputed Jacobians using a high-performance computing cluster, supporting potential real-time deployment. In contrast, *fused diffEIT* stacks impedance data from all probe placements into a single inverse problem, maximizing data redundancy and increasing voxel density. This integration of measurements with differing depth-sensitivity profiles explains the depth-dependence reversal observed in Figure 5: deeper voxels in *fused diffEIT* receive contributions from more measurements, preserving conductivity range with depth, whereas single-method reconstructions lose sensitivity. However, there are computational costs as the stacked data and 6 GB composite Jacobian extends processing time to 47.1 s along with reduced conductivity contrast, which alternative regularization or absolute-EIT formulations may mitigate. The current software is not optimized for real-time use and reconstructions times could be improved with modifications.

The SR and averaged methods use precomputed single-site reconstructions as inputs and increase pixel count tenfold while adding only 0.56 s and 0.05 s, respectively, to single-site reconstruction time. SR can sharpen boundaries when inputs are accurate, but exhibited increased FWHM and 10–90% width in experimental data, likely reflecting amplification of noise and artifacts present in the experimental single-site inputs. The averaged method produced more accurate boundary localization but more diffuse boundaries than SR.

Robustness was evaluated under additive Gaussian noise, simulated poor electrode contact, and EM localization error. All reconstruction methods maintained boundary localization within 0.5 mm across all noise sources. Notably, focal artifacts emerged in simulated reconstructions as electrodes were removed in which conductivity beneath removed electrodes progressively approaching saline values. These artifacts closely resemble the non-linear features observed in experimental reconstructions, suggesting suboptimal electrode contact in some regions of the phantom. *Fused diffEIT* showed minimal sensitivity to EM localization error, while SR and averaged methods exhibited changes once positional error reached 0.5 mm or rotational error exceeded 5%. Across all three noise sources, *fused diffEIT* demonstrated the greatest immunity.

Non-linear boundary distortions in the experimental data are consistent with poor contact in localized regions, corroborated by ~11% average IIVV pattern removal across the twelve measurements. The current pipeline retains only patterns valid across all measurements, removing ~25% of total data, though nearly 70,000 patterns remained available. Simulated removal at this level produced only minor changes in *fused diffEIT*. Allowing measurement-specific pattern exclusion in future implementations could recover this loss.

The approach presented here creates reconstructions with voxels 0.05–0.3 mm in size, overcoming the spatial limitations of Hu et al.'s (2022) 13 mm electrode array through higher electrode density and fusion-based data redundancy. Compared with Murphy et al.'s (2020) prostate biopsy probe, from which the *fused diffEIT* formulation is adapted, the present device makes solid surface contact and is maneuverable within the oral cavity, while EM tracking provides accurate localization without line-of-sight constraints.

This study is a feasibility demonstration, and several limitations should be noted. Reconstruction parameters of depth, area, number of measurement sites, and grid resolution were chosen empirically to establish proof of concept rather than to define performance limits. The agar phantom captures the >3× conductivity contrast between cancerous and healthy oral mucosa reported by Lloyd et al. (2025), but at higher absolute conductivities and with greater regional homogeneity than tissue; phantom-saline diffusion during the <10 min acquisition window may also alter regional conductivities. Reconstructions extend only to 1.2 mm depth, well below the 5 mm clinical margin definition, though the depth sensitivity of *fused diffEIT* motivates investigation of deeper mapping. Future work will incorporate more complex boundaries and ex vivo tissue to assess heterogeneity and deformation effects.

Two priorities for systematic characterization follow. First, the relationship between measurement count, reconstruction area, and achievable resolution must be quantified to identify the minimum measurements needed for reliable mapping over clinically relevant areas. Second, work systematically characterizing the maximum effective spatial resolution and voxel density could be performed using finer FEM meshes, different techniques for coarse to fine mapping, and spatial resolution phantoms for validation. Finally, regularization techniques beyond Laplace smoothing could be employed to optimize for sharp boundary edges in reconstructions.

By fusing spatially registered EIT measurements from an EM-tracked handheld SMA probe, this work demonstrates up to a ten-fold increase in voxel count over single-site reconstructions. *Fused diffEIT* offers the most favorable balance of accuracy, robustness, and with future optimization, computational feasibility that positions the method as a promising candidate for further development toward real-time intraoperative margin assessment.

## 5. Conclusions

This study demonstrates, for the first time, that spatially registered EIT measurements from an EM-tracked handheld probe can be fused to produce conductivity reconstructions with three-fold greater imaging area and ten-fold higher spatial resolution than single-site measurements. Four reconstruction approaches were evaluated through simulation, noise analysis, and phantom experiments, all achieving strong classification (AUC > 0.9). The fused difference-EIT method demonstrated the greatest robustness to noise, poor electrode contact, and EM localization error, while post-reconstruction fusion methods offered favorable trade-offs between resolution, computational cost, and boundary sharpness. These results establish fused EIT as a viable framework for extending the spatial coverage of compact impedance devices beyond their electrode footprint. This approach shows promise converting point-wise measurements into spatially resolved conductivity maps to address a critical limitation of current intraoperative margin assessment. Future work will focus on ex vivo tissue validation, extension to clinically relevant depths, and determination of the minimum probe poses required for efficient margin mapping.

## Acknowledgements

This work was supported by the National Science Foundation Research Traineeship, Transformative Research and Graduate Education in Sensor Science, Technology and Innovation (DGE- 2125733).